\documentclass[twocolumn,
superscriptaddress,
showpacs,preprintnumbers,
 amsmath,
 aps,
prb,
]{revtex4-1}
\usepackage{graphicx}
\usepackage{dcolumn}
\usepackage{bm}
\usepackage{color}

\newcommand{\beq}{\begin{equation}}
\newcommand{\eeq}{\end{equation}}
\newcommand{\beqa}{\begin{eqnarray}}
\newcommand{\eeqa}{\end{eqnarray}}

\begin{document}

\title{Fast spin control in a two-electron double quantum dot by dynamical
invariants}

\author{Yue Ban}
\affiliation{Departamento de Qu\'{\i}mica-F\'{\i}sica, UPV/EHU, Apdo 644, 48080 Bilbao, Spain}

\author{Xi Chen}
\affiliation{Department of Physics, Shanghai University, 200444 Shanghai, People's Republic of China}
\affiliation{Departamento de Qu\'{\i}mica-F\'{\i}sica, UPV/EHU, Apdo 644, 48080 Bilbao, Spain}

\author{J. G. Muga}
\affiliation{Departamento de Qu\'{\i}mica-F\'{\i}sica, UPV/EHU, Apdo 644, 48080 Bilbao, Spain}
\affiliation{Department of Physics, Shanghai University, 200444 Shanghai, People's Republic of China}

\author{E. Ya Sherman}
\affiliation{Departamento de Qu\'{\i}mica-F\'{\i}sica, UPV/EHU, Apdo 644, 48080 Bilbao, Spain}
\affiliation{IKERBASQUE, Basque Foundation for Science, 48011 Bilbao, Spain}

\begin{abstract}
Inverse engineering of electric fields has been recently proposed to achieve
fast and robust spin control in a single-electron quantum dot with
spin-orbit coupling. In this paper we design, by inverse engineering based
on Lewis-Riesenfeld invariants, time-dependent electric fields to realize
fast transitions in the selected singlet-triplet subspace of a two-electron
double quantum dot. We apply two-mode driving schemes, directly employing
the Lewis-Riesenfeld phases, to minimize the electric field necessary to
design flexible protocols and perform spin manipulation on the chosen
timescale.
\end{abstract}


\maketitle

\section{Introduction}

Coherent manipulation of quantum systems with time-dependent fields is a
major goal in different areas, including atomic, molecular, and optical
physics, as well as in semiconductor-based devices \cite
{Allen,Bergmann,Vitanov-Rev1,Kral,Molmer}, with applications in metrology,
interferometry, and quantum information processing. In all these areas,
control schemes for fast state transitions are highly desirable, and
techniques to design ``shortcuts to adiabaticity" \cite
{Rice,Berry09,Chen10a,Chen10b,ChenPRA,transport,opttransport,Sara,Masuda}
have been proposed to speed up slow processes and avoid decoherence effects
(see recent review \cite{review}).

Recent advances in device fabrication and measurement at the nanoscale are
approaching the goal of coherent and reliable manipulation of electron spins
in quantum dots (QDs) for quantum information processing \cite
{qubit,QD,Marcus2013,reviewDLoss}. Shortening the operation times is a major
challenge, not only to achieve fast computations, but also to avoid
decoherence. In a recent publication \cite{spinQD}, we have applied
shortcut-to-adiabaticity techniques, specifically invariant-based inverse
engineering \cite{Chen10a,ChenPRA}, to design time-dependent electric fields
that speed up spin flips in a two-dimensional (2D) ``lateral'' QD. Electric
control is made possible in the presence of spin-orbit (SO) coupling \cite
{RashbaEfros,spin resonance,Nowack,Loss}, and it is a promising alternative
to spin manipulation by magnetic spin resonance. A time-dependent electric
field can in principle be generated on the nanoscale by local electrodes
and, thus, provides individual access and efficient manipulation for each
spin \cite{Nowack}. By contrast, oscillating magnetic fields are not easy to
generate and manipulate locally.

Beyond the single dot, two-electron double quantum dots (DQDs) offer a
minimal basic frame for one and two-qubit gates \cite{Hanson,Petta}, and
alternative qubit encoding and control approaches. In this paper, we apply
invariant-based inverse engineering to design the electric fields necessary
to perform fast transitions in a singlet-triplet two-level subspace for a
lateral two-electron DQD with spin-orbit coupling. Significant differences
are found with respect to the single dot \cite{spinQD}, due to the new
structural dependence of the effective Hamiltonian with the applied fields.
This dependence in fact facilitates richer control possibilities for the
DQD. We develop a new method for invariant-based inverse engineering: a
multi-mode driving, first proposed in Ref. [\onlinecite{Chen3level}], that
uses all eigenstates of the dynamical invariant rather than only one of them
\cite{ChenPRA,spinQD}. Several examples of the control possibilities are
also provided.

\section{Model and Hamiltonian for spin driving}

\begin{figure}[tbp]
\scalebox{0.50}[0.50]{\includegraphics{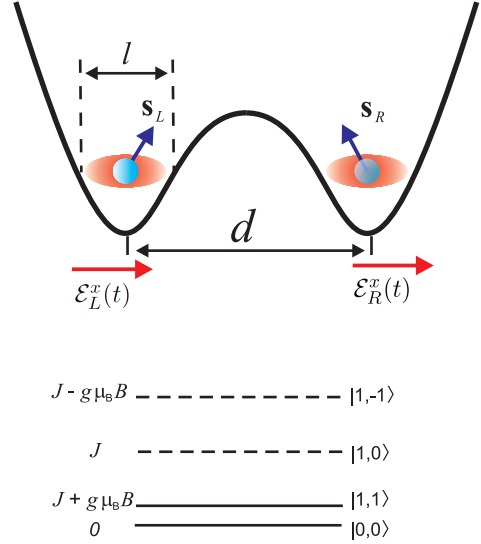}}
\caption{(Color online) (Upper panel) Schematic plot of electrons and applied electric
fields in a double quantum dot. (Lower panel) Spin states in the absence of external driving.}
\label{double-QD}
\end{figure}

We consider two electrons in a DQD formed in the $\left( x,y\right) $ plane
of a two-dimensional electron gas confined in the $z$-direction (see Fig. \ref{double-QD}, upper panel) in the presence of SO coupling, external static
magnetic field \cite{Coulomb1}, and an in-plane time-dependent electric
field. The Hamiltonian of this system has the following form:
\begin{equation}
H=H_{0}+H_{Z}+H_{\rm so}+V(t).
\end{equation}
The spin-independent Hamiltonian $H_{0}$ is:
\begin{equation}
H_{0}=\sum_{j=1,2}\left( \frac{{\mathbf p}_{j}^{2}}{2m}+U({\mathbf r}_{j})\right)
+\frac{e^{2}}{\epsilon \left| {\mathbf r}_{1}-{\mathbf r}_{2}\right| },
\end{equation}
where index $j$ numerates the electrons with corresponding coordinates, $\epsilon$
is the dielectric constant, and $m$ is the electron effective mass. The
potential $U({\mathbf r}_{j})$ describes the confinement and has two minima, separated by
the distance $d$ in the $(x,y)$-plane, with the electron wavefunctions
well-localized near one minimum on the spatial scale $l\ll d/2$.
We assumed that the magnetic field is weak and thus neglected the
contribution of the corresponding vector potential in the momentum for
electrons. In the limit of small overlap, the eigenstates of $H_{0}$ can be
accurately presented \cite{Fazekas} in the symmetrized
form $\left( \psi _{L}({\mathbf r}_{1})\psi _{R}({\mathbf r}_{2})\pm
\psi _{L}({\mathbf r}_{2})\psi _{R}({\mathbf r}_{1})\right) \left| S,S_{z}\right\rangle$, where
$L$ (left) and $R$ (right) correspond to the position of the minimum, $S$
and $S_{z}$ are the total spin and its $z-$component, respectively, and the
sign in the brackets is determined by fermionic permutation law for the given $S$.
Finally, $H_{0}$  can be presented as the product of site-related spin
operators $H_{0}=J(\bm{s}_{L}\cdot\bm{s}_{R})$, where $J$ is the
corresponding exchange integral \cite{Fazekas}, and $\bm{s}_{L,R}$ is the total spin
(produced by the two identical electrons) located near the corresponding
minimum.

The Zeeman term for magnetic field $B_{z}\left(x,y\right)\parallel z-$axis
has the form
\begin{equation}
H_{Z}=\Delta \left( x_{1},y_{1}\right) s_{1}^{z}+\Delta \left(
x_{2},y_{2}\right) s_{2}^{z},
\end{equation}
with $\Delta\left(x_{j},y_{j}\right) =\mu_{B}gB_{z}\left(x_{j},y_{j}\right)$,
Bohr magneton $\mu_{B},$ and the Land\'{e} factor $g$. Although the magnetic field can
be strongly inhomogeneous, we assume that it is uniform inside the dot on the
spatial scale of the order of $l$ and obtain for the well-localized
electrons:
\begin{equation}
H_{Z}=\Delta _{L}s_{L}^{z}+\Delta _{R}s_{R}^{z}.
\label{HZ}
\end{equation}
We choose the Hamiltonian of SO coupling $H_{\text{so}}$ in the
form
\begin{equation}
H_{\text{so}}=\sum_{j}\alpha (\sigma _{j}^{x}p_{j}^{y}-\sigma
_{j}^{y}p_{j}^{x})+\sum_{j}\beta \sigma _{j}^{z}p_{j}^{x},
\end{equation}
to describe the structure-related Rashba and bulk-originated Dresselhaus SO
coupling for the assumed $[110]$ growth axis \cite{beta-110growth} with the
interaction parameters $\alpha $ and $\beta $, respectively. It is
well-known, that in the quantum dots, where electrons are localized, the
direct role of SO coupling is very weak and can be neglected.

However, $H_{\text{so}}$ is important for the spin driving by electric
field (see e.g. [\onlinecite{Golovach,Nowak}] and references therein). Here this approach will be
applied to develop a protocol of controllable spin driving by electric field in a
two-electron DQD. For this purpose we present
\begin{equation}
V(t)=-\frac{1}{2}\frac{e}{c}\left[ \left\{ \mathbf{A(}x_{1,}y_{1}),\mathbf{v}%
_{1}\right\} +\left\{ \mathbf{A(}x_{2,}y_{2}),\mathbf{v}_{2}\right\} \right],
\end{equation}
where $\mathbf{A(}x_{j},y_{j})$ is the vector potential of time-dependent
electric field ${\bm{\mathcal{E}}}_{j}(t)=-(1/c)\partial \mathbf{A}_{j}/\partial t$,
and the bracket $\left\{,\right\}$ stands for the anticommutator.
In the following calculation, we shall neglect $H_\text{so}$-term coupling directly to the
electron momentum assuming that the transitions between orbital states can be
disregarded. More important, as a result of SO coupling, $x$ and $y$%
-axis components of velocity $\mathbf{v}$, such as, for example, $%
v_{x}=i\left[ H_0 + H_{\text{so}}, x\right]/\hbar$, acquire spin-dependent contribution,
stemming from the commutators of the corresponding coordinate with the
linear momentum in $H_{\text{so}}$-term. Then, the coupling
acquires the form:
\begin{equation}
V(t)=-2\frac{e}{c}\sum_{j}\left[ {A}_{j}^{x}\mathbf{(}x_{j,}y_{j})\left(
\beta s_{j}^{z}-\alpha s_{j}^{y}\right) +{A}_{j}^{y}\mathbf{(}%
x_{j,}y_{j})\alpha s_{j}^{x}\right] ,
\end{equation}
very similar to that for the magnetic field. Again, we consider vector
potential uniform inside the dots and get for localized states
\begin{eqnarray}
V(t)&=& -2\frac{e}{c}\left[{A}_{L}^{x}\left( \beta s_{L}^{z}-\alpha
s_{L}^{y}\right) +{A}_{L}^{y}\alpha s_{L}^{x}\right. \nonumber \\
&&+\left.{A}_{R}^{x}\left( \beta s_{R}^{z}-\alpha s_{R}^{y}\right) +{A}_{R}^{y}\alpha
s_{R}^{x}\right],
\end{eqnarray}
where we consider spin components per dot similar to Eq. (\ref{HZ}) and omit
explicit $(x,y)-$dependence of $\mathbf{A}$ in the formulas.

To simplify the consideration further, we set the same magnetic field
$B$ for both dots and introduce $\Delta\equiv\Delta_{L}=\Delta _{R}=\mu
_{B}gB$. In this static field the four eigenstates of this system can be
expressed by singlet and triplets for total spin $S=0$ and $S=1$, respectively.
We assume that the energy difference between the singlet $|0,0\rangle$ and
the lowest one of the triplets $|1,1\rangle$ (for typical $g<0$) is
much less than $J$ that is $|J+\Delta |\ll J$. Here we concentrate on
transitions between these two states (see Fig. \ref{double-QD}, lower panel).

In the basis $|0,0\rangle\equiv|1\rangle=(1,0)^{T}$ and $|1,1\rangle\equiv|2\rangle=(0,1)^{T}$,
where the $T$ stands for transpose,
the total spin-dependent Hamiltonian (neglecting $H_{\text{so}}$-term) becomes 
\begin{equation}
H=\frac{\hbar }{2}\left(
\begin{array}{cc}
Z_{1} & X+iY \\
X-iY & Z_{2}
\end{array}
\right),
\label{Htotal}
\end{equation}
where the elements of the matrix are:
\begin{eqnarray}
\label{X} X &=&\frac{\sqrt{2}\alpha }{\hbar }\frac{e}{c}(A_{L}^{y}-A_{R}^{y}),
\label{Z2_H} \\
Y &=&-\frac{\sqrt{2}\alpha }{\hbar }\frac{e}{c}(A_{L}^{x}-A_{R}^{x}), \\
Z_{1} &=&-\frac{3J}{2\hbar }, \\
Z_{2} &=&\frac{1}{\hbar }\left[ \frac{J}{2}+2\Delta -2\beta \frac{e}{c}%
(A_{L}^{x}+A_{R}^{x})\right] .
\end{eqnarray}
%
\begin{figure}[tbp]
\scalebox{0.40}[0.40]{\includegraphics{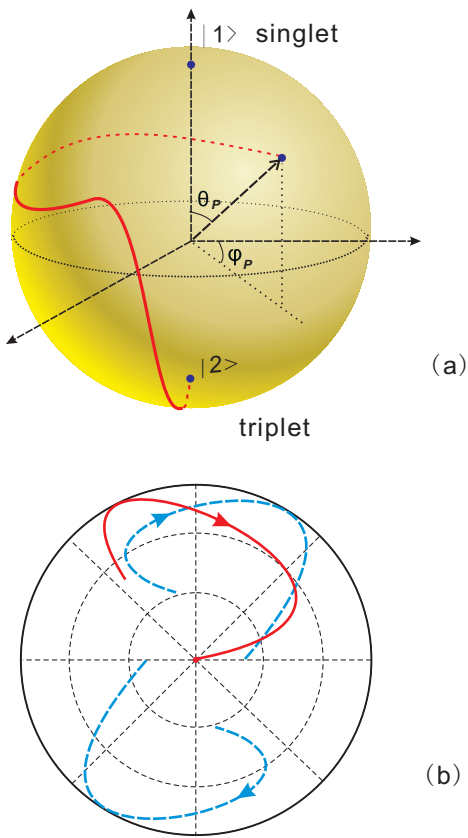}}
\caption{(Color online) (a) Schematic plot of spin control from an arbitrary
state to the target state $|2{\rangle }$ on the Bloch sphere, where the
north and south poles represent the singlet state $|1{\rangle }=|0,0{\rangle
}$ and triplet state $|2{\rangle }=|1,1{\rangle }$. (b) Schematic projection
from top view on the $x$-$y$ plane of the auxiliary paths of two eigenstates
of the dynamical invariant (dashed blue line), compared to the physical path
of state evolution driven by the designed Hamiltonian (solid red line). }
\label{model}
\end{figure}

Here, unlike the single QD \cite{spinQD}, counterdiabatic protocol (or
quantum transitionless driving) \cite{Rice,Berry09,Chen10b} could be applied
for electric spin control\cite{Yue}. However, rather than
relying on a potentially complicated control of required four parameters, we restrict
ourselves to the simplifying assumptions $A_{L}^{y}=A_{R}^{y}=X=0$, which
leaves the electric field $x$-components the only controlled function. The
gauge is fixed by assuming ${\mathcal{E}}_{L,R}^{x}=0$ for all $t<0$ and $%
A_{L,R}^{x}(0)=0$ so that the electric fields start to be built up
from $t=0$. In the following sections we inversely engineer
the time dependence of required electric fields for arbitrary operations
by using corresponding invariants of the motion.


\section{Dynamical invariants and inverse engineering}

For completeness we shall first briefly review the Lewis-Riesenfeld
invariant theory applied to a two-level system. Specifically for our DQD
system we shall then design by two-mode inverse engineering the electric
fields to induce a particular transition. A dynamical invariant $I$ of the
Hamiltonian $H$ should satisfy
\begin{eqnarray}  \label{invariant-condition}
\frac{d I(t)}{dt} \equiv \frac{\partial I(t)}{\partial t} + \frac{i}{\hbar} [%
H(t),I(t)] = 0,
\end{eqnarray}
so that its expectation values remain constant in time. Parameterizing the
Bloch sphere by a polar angle $\theta_a (t) \equiv \theta_a$ and the
azimuthal angle $\varphi_a (t) \equiv \varphi_a$, we may express the
orthogonal eigenstates $|\chi_{\pm} (t) {\rangle}$ of the invariant $I(t)$
as 
\begin{eqnarray}
|\chi_{+}(t) {\rangle} &=& \left(
\begin{array}{c}
\cos\displaystyle{\frac{\theta_a}{2}} e^{i \varphi_a} \\
\sin\displaystyle{\frac{\theta_a}{2}}
\end{array}
\right), \\
|\chi_{-}(t) {\rangle} &=& \left(
\begin{array}{c}
\sin \displaystyle{\frac{\theta_a}{2}} \\
- \cos \displaystyle{\frac{\theta_a}{2}} e^{-i \varphi_a}
\end{array}
\right).
\end{eqnarray}
Assuming that $I(t) |\chi_\pm (t){\rangle} = \lambda_\pm |\chi_\pm (t){%
\rangle}$ with eigenvalues $\lambda_{\pm}=\pm g \mu_B B_{c}/2$, we can write
the invariant as \cite{ChenPRA}
\begin{eqnarray}  \label{invariant}
I (t) = \frac{g \mu_B }{2} B_{c} \left(
\begin{array}{cc}
\cos{\theta_a} & \sin{\theta_a} e^{i \varphi_a} \\
\sin{\theta_a} e^{-i \varphi_a} & -\cos{\theta_a}
\end{array}
\right),
\end{eqnarray}
where $B_{c}$ is an arbitrary constant magnetic field to keep $I(t)$ with
units of energy. According to Lewis-Riesenfeld theory, the solution of the
Schr\"{o}dinger equation, $i \hbar \partial_t \Psi=H(t) \Psi$, is a
superposition of orthonormal ``dynamical modes" \cite{LR}, $\Psi (t) =
\sum_n c_n e^{i \gamma_n (t)} |\chi_{n} (t) {\rangle}$ ($n=\pm$), where $c_n$
are time-independent amplitudes and $\gamma_n (t)$ are Lewis-Riesenfeld
phases,
\begin{equation}  \label{LR-phases}
\gamma_n \equiv \gamma_n (t) =\frac{1}{\hbar} \int^t_0 \langle
\chi_n(t^{\prime}) | i\hbar \frac{\partial }{\partial t^{\prime}} -
H(t^{\prime})| \chi_n(t^{\prime}) \rangle dt^{\prime}.
\end{equation}
Substituting Eqs. (\ref{Htotal}) and (\ref{invariant}) into Eq. (\ref
{invariant-condition}) and combining it with Eq. (\ref{LR-phases}), we may
write the Hamiltonian matrix elements in terms of invariant eigenvector
angles as
\begin{eqnarray}  \label{Z2}
X &=& (\dot{\gamma}_- - \dot{\gamma}_+ - \dot{\varphi}_a) \sin\theta_a
\cos\varphi_a + \dot\theta_a \sin\varphi_a, ~~ \\
Y &=& (\dot{\gamma}_- - \dot{\gamma}_+ - \dot{\varphi}_a) \sin\theta_a
\sin\varphi_a - \dot\theta_a \cos\varphi_a, ~~ \\
Z_1 &=& -(\dot{\gamma}_- + \dot{\gamma}_+) - \dot{\varphi}_a + (\dot{\gamma}%
_- - \dot{\gamma}_+ - \dot{\varphi}_a) \cos\theta_a,~~~ \\
Z_2 &=& -(\dot{\gamma}_- + \dot{\gamma}_+) + \dot{\varphi}_a - (\dot{\gamma}%
_- - \dot{\gamma}_+ - \dot{\varphi}_a) \cos\theta_a,~~~
\end{eqnarray}
from which Lewis-Riesenfeld phases are found to obey
\begin{eqnarray}  \label{auxiliary equations4}
\dot{\gamma}_+ &=& \frac{3 J}{4 \hbar} + \frac{1}{2} \dot{\theta}_a \tan{%
\frac{\theta_a}{2}} \tan\varphi_a -\dot{\varphi}_a, \\
\dot{\gamma}_- &=& \frac{3 J}{4 \hbar} - \frac{1}{2} \dot{\theta}_a \cot{%
\frac{\theta_a}{2}} \tan\varphi_a,
\end{eqnarray}
and the vector potential components take the form
\begin{eqnarray}  \label{AxR}
A^x_L &=& \frac{\hbar c}{2 \beta e} \left[ \frac{J+\Delta}{\hbar} - \dot{%
\varphi}_a \right. \\
&&- \left. \dot{\theta}_a \cot{\theta_a}\tan{\varphi_a} + \frac{\beta \dot{%
\theta}_a}{\sqrt{2}\alpha} \sec{\varphi_a} \right],  \nonumber \\
A^x_R &=& \frac{\hbar c}{2 \beta e} \left[ \frac{J+\Delta}{\hbar} - \dot{%
\varphi}_a\right. \\
&&- \left. \dot{\theta}_a \cot{\theta_a}\tan{\varphi_a} -\frac{\beta \dot{%
\theta}_a}{\sqrt{2}\alpha} \sec{\varphi_a} \right],  \nonumber
\end{eqnarray}
when $A^y_{L} = A^y_{R} =0$ is imposed. Note the possibility of finding
divergences when $\varphi_a$ is an odd multiple of $\pi/2$ and when $%
\theta_a $ is a multiple of $\pi$.

Once the functions $\theta _{a}(t)$ and $\varphi _{a}(t)$ are known, the
electric fields, ${\mathcal{E}}_{L}^{x}(t)=-(1/c)\partial A_{L}^{x}/\partial
t$ and ${\mathcal{E}}_{R}^{x}(t)=-(1/c)\partial A_{R}^{x}/\partial t$, can
be calculated. As we keep $A_{L,R}^{x}(0)=0$, the following two constraints
should hold:
\begin{eqnarray}
\label{limitation-theta}
\dot{\theta}_{a}(0)&=&0,
\\
\label{limitation-varphi}
\dot{\varphi}_{a}(0)&=&\frac{J+\Delta }{\hbar }.
\end{eqnarray}
Next, we shall discuss several examples.

\section{Transfer an arbitrary initial state to the target one}

Suppose that we want to transfer the arbitrary initial state at $t=0$,
\begin{eqnarray}
\Psi(0)= \left(
\begin{array}{c}
\cos{\displaystyle{\frac{\theta_p (0)}{2}}} e^{i \varphi_p (0)} \\
\sin{\displaystyle{\frac{\theta_p (0)}{2}}}
\end{array}
\right),
\end{eqnarray}
to $\Psi(t_f)=|2{\rangle}$ as the target state at the final time $t=t_f$.
The polar and azimuthal angles $\theta_p \equiv \theta_p(t)$ and $\varphi_p
\equiv \varphi_p(t)$, represent the dynamical path on the Bloch sphere of
the two-level system, see Fig. \ref{model}. In general,
\begin{equation}  \label{wavefunction}
\Psi(t) = c_{+} e^{i \gamma_{+} (t)} |\chi_{+} (t) {\rangle} + c_{-} e^{i
\gamma_{-} (t)} |\chi_{-} (t) {\rangle},
\end{equation}
where the time-independent coefficients $c_{\pm} = {\langle} \chi_{\pm} (0)
| \Psi (0){\rangle}$ are given by %
\begin{eqnarray}
\label{c+}
c_{+} &=& \cos{\frac{\theta_a (0)}{2}} \cos{\frac{\theta_p (0)}{2}} e^{i[\varphi_p(0)-\varphi_a (0)]} \\
&& + \sin{\frac{\theta_a (0)}{2}} \sin{\frac{\theta_p (0)}{2}},  \nonumber \\
\label{c-}
c_{-} &=& \sin{\frac{\theta_a (0)}{2}} \cos{\frac{\theta_p (0)}{2}} e^{i
\varphi_p(0)}   \\
&&- \cos{\frac{\theta_a (0)}{2}} \sin{\frac{\theta_p (0)}{2}} e^{i \varphi_a
(0)}.  \nonumber
\end{eqnarray}
%
The time-dependent populations $P_1(t) = |{\langle} 1| \Psi (t) {\rangle}|^2$
and $P_2(t) = |{\langle} 2| \Psi(t) {\rangle}|^2=1-P_1(t)$, can thus be
explicitly expressed as
\begin{eqnarray}
\label{wavefunction-1}
P_1(t) &=& \left|c_+ \cos\frac{\theta_a}{2} e^{i (\varphi_a + \gamma_+)} +
c_- \sin\frac{\theta_a}{2} e^{i \gamma_-}\right|^2, \\
\label{wavefunction-2} P_2(t) &=& \left|c_+ \sin\frac{\theta_a}{2} e^{i \gamma_+} - c_- \cos\frac{%
\theta_a}{2} e^{-i (\varphi_a- \gamma_{-})} \right|^2.
\end{eqnarray}
To reach the final state $\Psi(t_f) = |2{\rangle}$, up to a phase factor,
the condition $P_1 (t_f) =0$ [$P_2(t_f) =1$] should be satisfied, so that we
can further set boundary conditions for $\theta_a (t_f)$. Once the boundary
conditions for $\theta_a$ and $\varphi_a$ are fixed, we interpolate $%
\theta_a(t)$ and $\varphi_a(t)$, and finally construct the Hamiltonian $H(t)$
and the time-dependent electric fields.

If, for simplicity, $\varphi_a(0)$ is set equal to the initial physical
angle $\varphi_p(0)$, the coefficients $c_\pm$ in Eqs. (\ref{c+}) and (\ref{c-}) are
\begin{eqnarray}  \label{coefficientc+}
c_+ &=& \cos\frac{\eta}{2}, \\
\label{coefficientc-}
c_- &=& e^{i \varphi_p(0)} \sin\frac{\eta}{2},
\end{eqnarray}
with $\eta = \theta_a(0) - \theta_p(0) $. On the Bloch sphere, $\Psi(0)$ and
$|\chi_+(0){\rangle}$ have the same longitude, while $|\chi_-(0){\rangle}$,
which is orthogonal to $|\chi_+(0){\rangle}$, possesses the azimuthal angle $%
\pi+\varphi_p(0)$. The condition for $P_2 =1$ [$P_1=0$] at the final time $%
t= t_f$, see Eqs. (\ref{wavefunction-1}) and (\ref{wavefunction-2}), is %
%
\begin{eqnarray}  \label{final P}
\hspace{-0.4cm}1 + \cos \theta_a(t_f) \cos \eta + \sin \theta_a(t_f) \sin
\eta \cos u(t_f) = 0,
\end{eqnarray}
%
where
\begin{eqnarray}  \label{ut}
u(t) = -\int^{t}_0 \dot{\theta}_f(\tau) \frac{\tan \varphi_a(\tau)}{\sin
\theta_a(\tau)} d\tau.
\end{eqnarray}
We have used
\begin{eqnarray}  \label{LR-difference}
\gamma_-(t) - \gamma_+(t) = \varphi_a(t) - \varphi_a(0) + u(t),
\end{eqnarray}
which follows from $X=0$, see Eq. (\ref{X}). We note from Eqs. (\ref{coefficientc+}) and (\ref{coefficientc-}) that $\eta$ determines the
relative weights of the two eigenstates. For two-mode driving, when $\eta$
is not equal to $0$ or $\pi$, Eq. (\ref{final P}) holds only when $\cos
u(t_f) = \pm 1$.

\subsection{Example 1: $\cos u(t_f) = 1$}

We choose first $\cos u(t_f) = 1$. In this case, $u(t_f) = 2 k \pi $ ($k=0,
\pm 1, \pm 2 ...$) and the condition (\ref{final P}) becomes
\begin{equation}
\cos[\theta_a (t_f) - \eta] =-1,
\end{equation}
which gives
\begin{eqnarray}  \label{thetat1}
\theta_a(t_f) = \theta_a(0) - \theta_p(0) + \pi.
\end{eqnarray}
To fulfill Eq. (\ref{thetat1}) and Eq. (\ref{limitation-theta}), we use the
second order polynomial \textit{Ansatz}
\begin{eqnarray}  \label{thetat-example1}
\theta_a(t) = \sum_{j=0}^2 a_j t^j,
\end{eqnarray}
where the coefficients $a_j$ can be found with a given $\theta_a(0)$. In
addition, we interpolate $\varphi_a$ by
\begin{eqnarray}  \label{varphit-example1}
\tan \varphi_a(t) = \left(\sum_{j=0}^2 b_j t^j\right) \sin \theta_a(t),
\end{eqnarray}
to avoid singularities in Eq. (\ref{ut}). Here $b_0 = \tan \varphi_p(0) /
\sin \theta_a(0)$ is obtained from $\varphi_a(0)=\varphi_p(0)$, and $b_1$
and $b_2$ can be solved from Eq. (\ref{limitation-varphi}) and $u(t_f)=2 k
\pi$.

\begin{figure}[]
\scalebox{0.70}[0.70]{\includegraphics{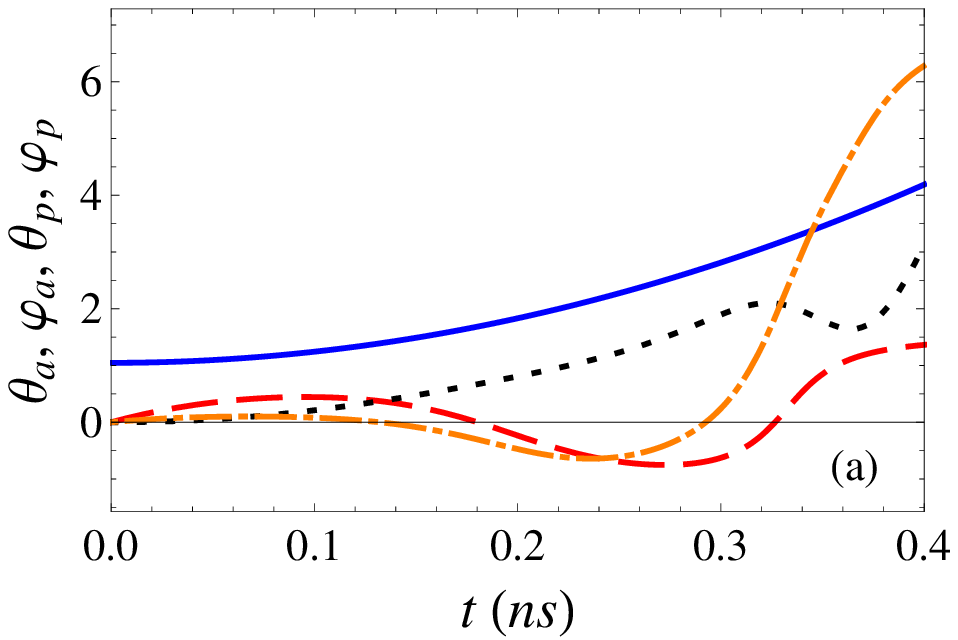}} \scalebox{0.70}[0.70]{%
\includegraphics{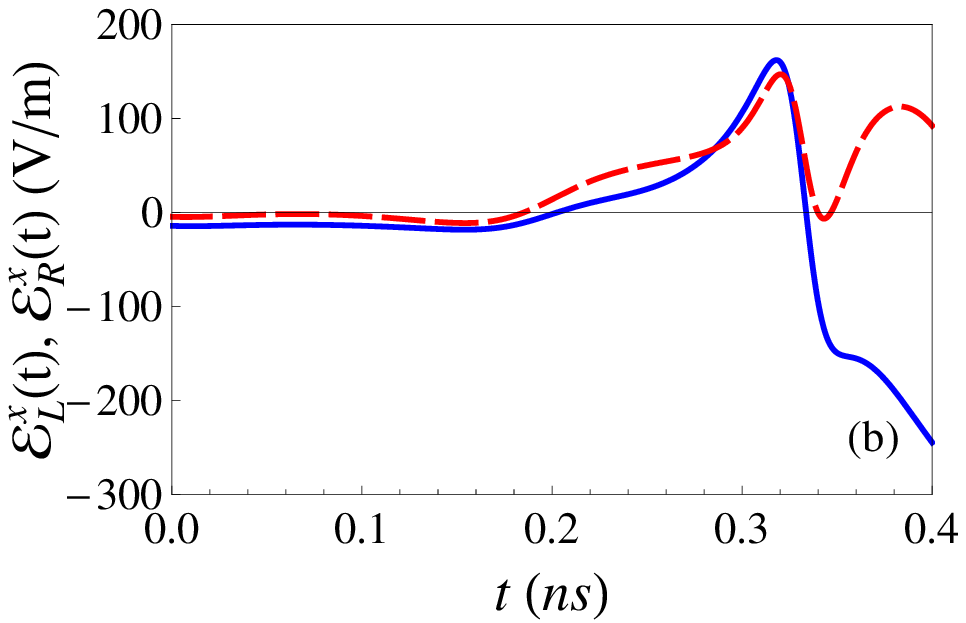}} \scalebox{0.70}[0.70]{%
\includegraphics{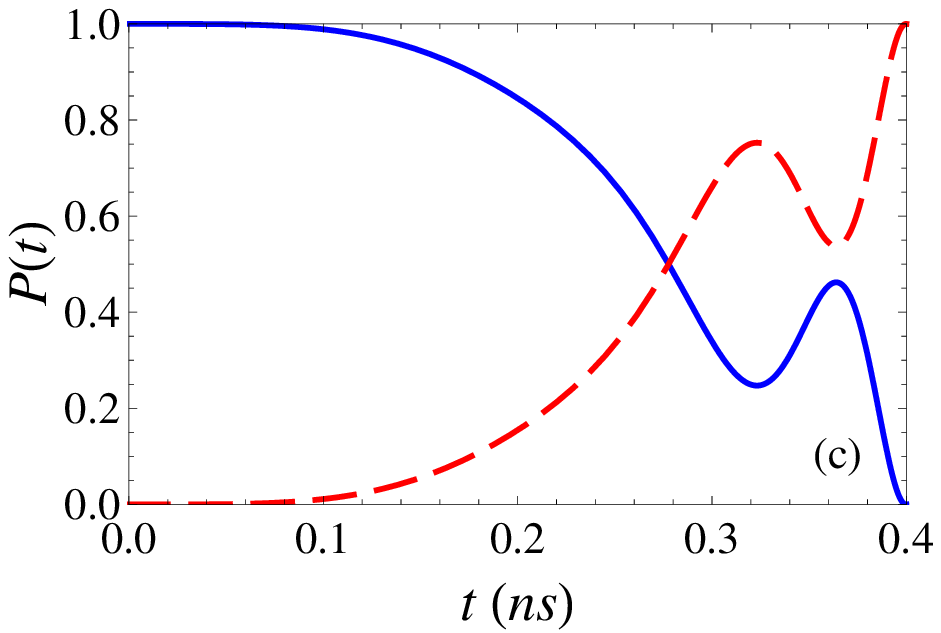}}
\caption{(Color online) Example 1 for two-mode driving with the initial
conditions: $\theta_p(0) = 0$, $\varphi_p(0) = 0$, $\theta_a(0)=\pi/3$, and $%
\varphi_a(0)=\varphi_p(0)$. For $u(t_f)=2\pi$: (a) The two auxiliary angles,
$\theta_a$ (solid blue line) and $\varphi_a$ (dashed red line), determined
by Eqs. (\ref{thetat-example1}) and (\ref{varphit-example1}), compared to
the physical polar angle $\theta_p$ (dotted black line) and the physical
azimuthal angle $\varphi_p$ (dot-dashed orange line). (b) The designed
electric fields ${\mathcal E}^x_L$ (solid blue line) and ${\mathcal E}^x_R$ (dashed red line).
(c) The populations $P_1$ (solid blue line) and $P_2$ (dashed red line).
Parameters: $t_f=0.4$ ns, $B_1=B_2=3.67$ T, $g=-0.44$, $J=0.1$ meV
(consequently, $|J + \Delta|/J = 0.06$), $\hbar \beta =0.25\times10^{-6}$ meV%
$\cdot$cm, and $\alpha = \beta /2$.}
\label{example1}
\end{figure}

Let us consider a population inversion or singlet-triplet transition from
the initial state $|1{\rangle} $ (we set $\theta_p(0)=0$ and $\varphi_p(0)=
0 $) to final state $|2 {\rangle}$ (up to a phase factor) for $t_f=0.4$ ns.
For this operation time, the relevant energy scale is below $0.01$ meV, much
less than the singlet-triplet spitting $J=0.1$ meV, such that transitions to
the two higher triplet states do not occur.

If we choose $\theta_a (0)=\pi/3$ and $k=1$, the functions of $\theta_a(t)$
and $\varphi_a(t)$ can be calculated from Eqs. (\ref{thetat-example1}) and (%
\ref{varphit-example1}), as shown in Fig. \ref{example1} (a). The
corresponding electric fields are depicted in Fig. \ref{example1} (b). The
population dynamics is numerically calculated by solving the time-dependent
Schr\"{o}dinger equation. Fig. \ref{example1} (c) demonstrates that the spin
state evolves from the initial state $|1{\rangle}$ to final one $|2{\rangle}$%
, up to a phase factor $\exp[i \varphi_p (t_f)]$, for the fixed time $t_f =
0.4$ ns. {\ The physical angles $\theta_p(t)$ and $\varphi_p(t)$
corresponding to the trajectory of the spin state on the Bloch sphere are
also shown in Fig. \ref{example1} (a).}

\begin{figure}[]
\scalebox{0.70}[0.70]{\includegraphics{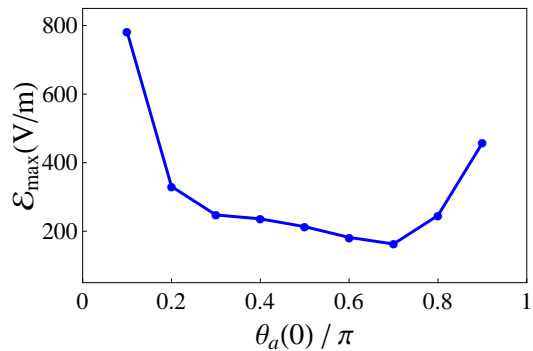}}
\caption{(Color online) Maximal value ${\mathcal E}_{\max}$ of the electric fields
versus the initial boundary condition $\theta_a (0)$ in the example 1, with
Eqs. (\ref{thetat-example1}) and (\ref{varphit-example1}). Other parameters
are the same as those in Fig. \ref{example1}. }
\label{Emax}
\end{figure}

The two-mode driving scheme provides flexibility to pick up different
boundary conditions for $\theta_a$ and $\varphi_a$, so as to minimize the
fields. To demonstrate this, we choose different initial values of $%
\theta_a(0)$. The maximal absolute value ${\mathcal E}_{\max}$ of applied electric
fields, ${\mathcal E}^x_L$ and ${\mathcal E}^x_R$, that is, ${\mathcal E}_{\max} = \max(|{\mathcal E}^x_L|,|{\mathcal E}^x_R|)$%
, can be decreased by choosing a suitable $\theta_a (0)$, as shown in Fig.
\ref{Emax}. Note that each two-mode scheme is equivalent to a single-mode
one, when the physical polar and azimuthal angles, $\theta_p$ and $\varphi_p$%
, are reinterpreted {\ by} the two auxiliary angles $\theta_a$ and $%
\varphi_a $ for designing the eigenstates of the invariant. However, the
functions of these angles are generally not simple.

%
\begin{figure}[ht]
\scalebox{0.70}[0.70]{\includegraphics{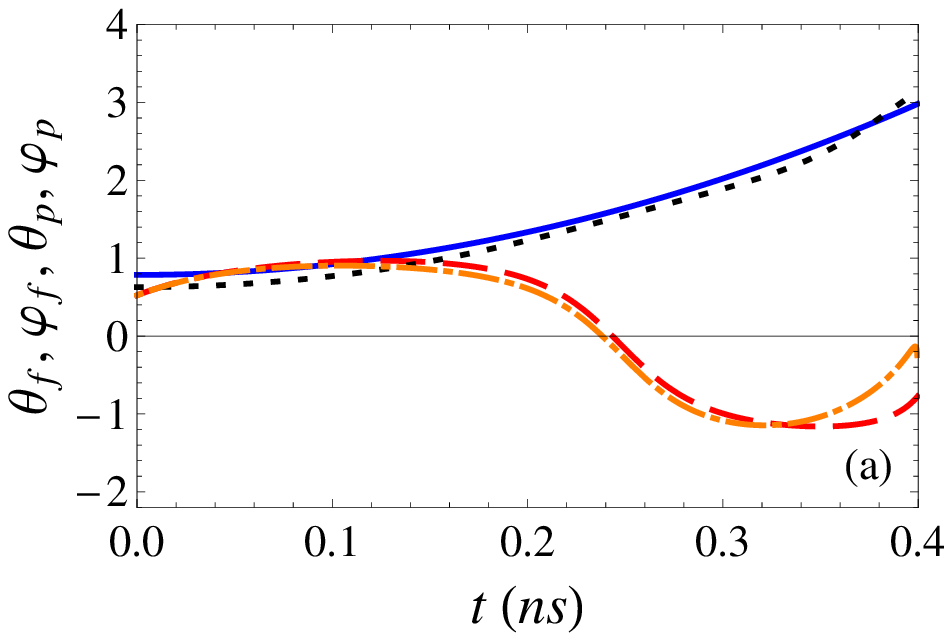}} \scalebox{0.70}[0.70]{%
\includegraphics{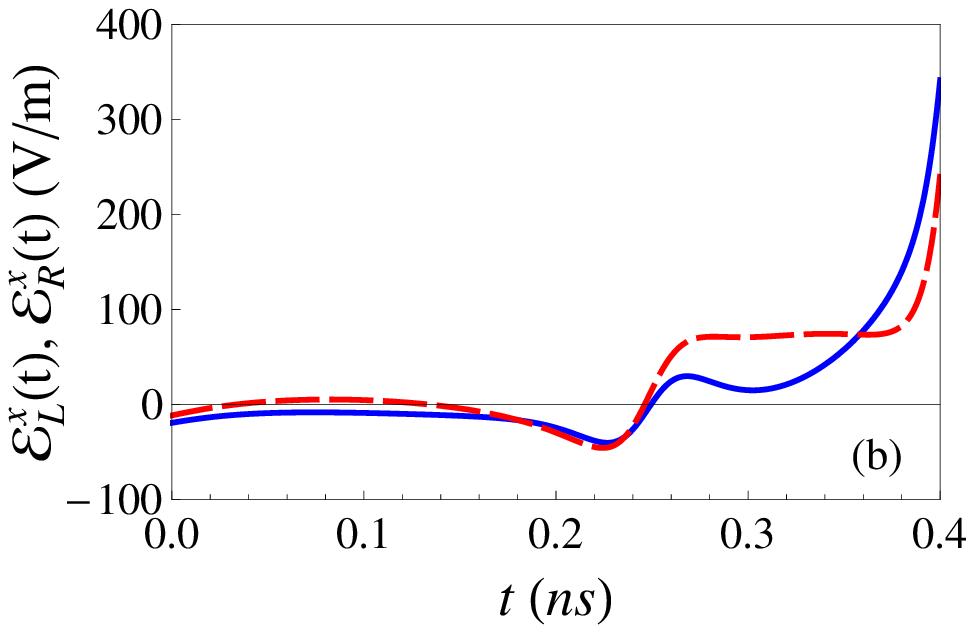}} \scalebox{0.70}[0.70]{%
\includegraphics{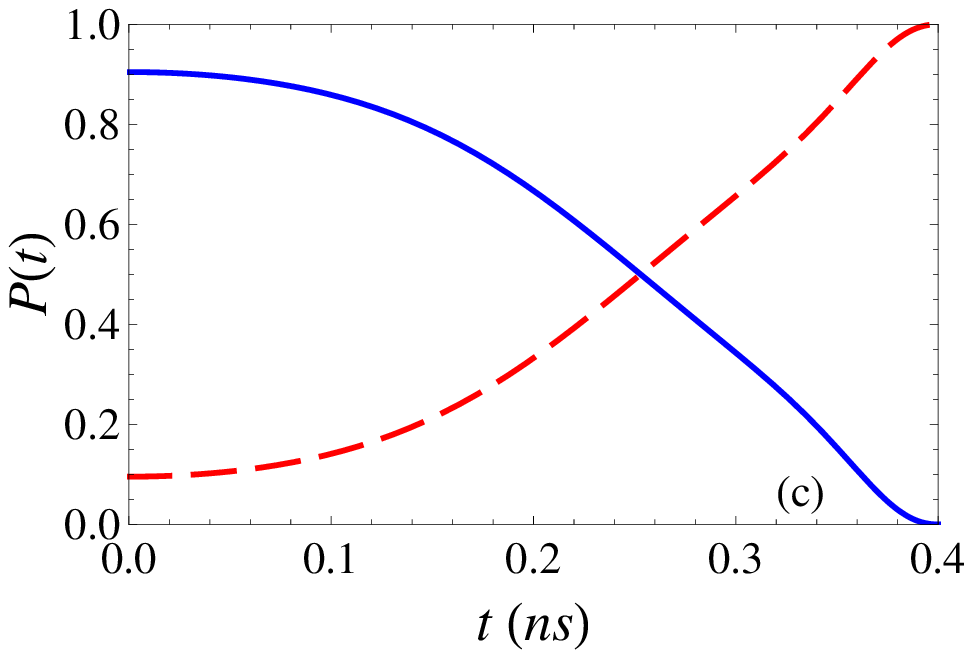}}
\caption{(Color online) Example 2 for two-mode driving with initial
conditions: $\theta_p(0) = \pi/5$, $\varphi_p(0) = \pi/6$, $%
\theta_a(0)=\pi/4 $, and $\varphi_a(0)=\varphi_p(0)$. For $u(t_f)=\pi$: (a)
The two auxiliary angles, $\theta_a$ (solid blue line) and $\varphi_a$
(dashed red line), determined by Eqs. (\ref{thetat-example1}) and (\ref
{varphit-example1}), compared to the physical polar angle $\theta_p$ (dotted
black line) and the physical azimuthal angle $\varphi_p$ (dot-dashed orange
line). (b) The designed electric fields ${\mathcal E}^x_L$ (solid blue line) and $%
{\mathcal E}^x_R$ (red dashed line). (c) The populations $P_1$ (solid blue line) and $%
P_2$ (dashed red line). Other parameters are the same as in Fig. \ref
{example1}. }
\label{example2}
\end{figure}
%

\subsection{Example 2: $\cos u(t_f) = -1$}

For the choice $\cos u(t_f) = -1$, we have $u(t_f) = (2k-1) \pi$ ($k= 0,
\pm1,\pm2 ...$) and the condition for $P_2(t_f) =1$ becomes
\begin{equation}
\cos[\theta_a (t_f) + \eta] = - 1,
\end{equation}
which results in
\begin{eqnarray}  \label{thetat2}
\theta_a(t_f) = -\theta_a(0) + \theta_p(0) + \pi.
\end{eqnarray}

In the example 2, we want to manipulate the spin state from an arbitrary
state to $|2 {\rangle}$ (up to a phase factor). We apply the same forms of $%
\theta_a$ and $\varphi_a$ as before [see Eqs. (\ref{thetat-example1}) and (%
\ref{varphit-example1})], and choose a different initial state $\theta_p(0)
= \pi/5$, $\varphi_p(0) = \pi/6$, and $\theta_a(0)=\pi/4$. For the fixed
time $t_f = 0.4$ ns, the two auxiliary angles $\theta_a$ and $\varphi_a$ are
determined by Eqs. (\ref{thetat-example1}) and (\ref{varphit-example1}), as
seen in Fig. \ref{example2} (a), where the physical angles of $\theta_p$ and
$\varphi_p$ are compared. The electric fields, ${\mathcal E}^x_L$ and ${\mathcal E}^x_R$ are
shown in Fig. \ref{example2} (b). The populations are represented in Fig.
\ref{example2} (c), in which the target state $|2{\rangle}$ is achieved at
final time $t_f =0.4$ ns.

\section{Transfer the initial state to an arbitrary state}

\begin{figure}[]
\scalebox{0.70}[0.70]{\includegraphics{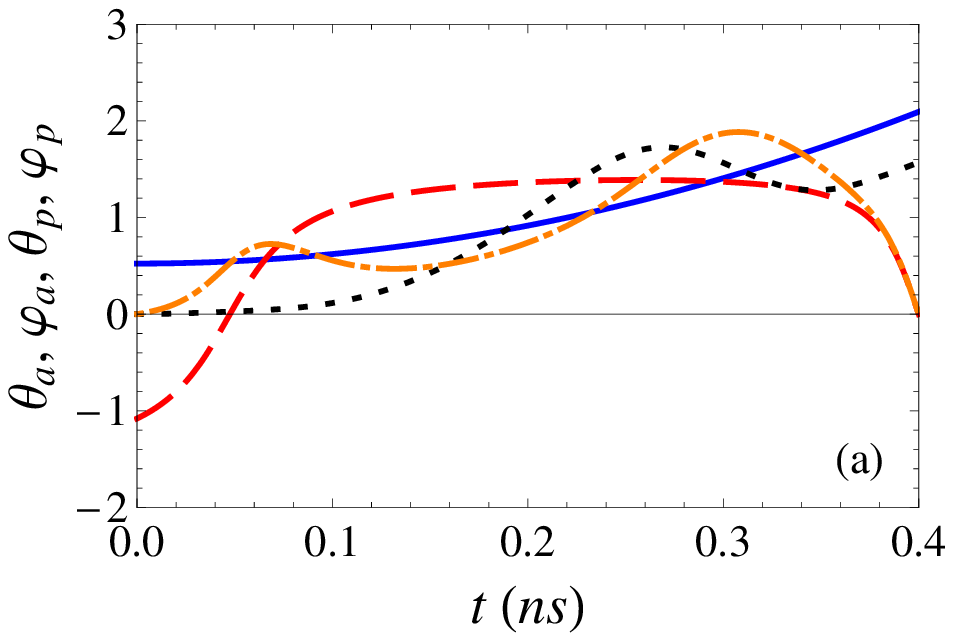}} \scalebox{0.70}[0.70]{%
\includegraphics{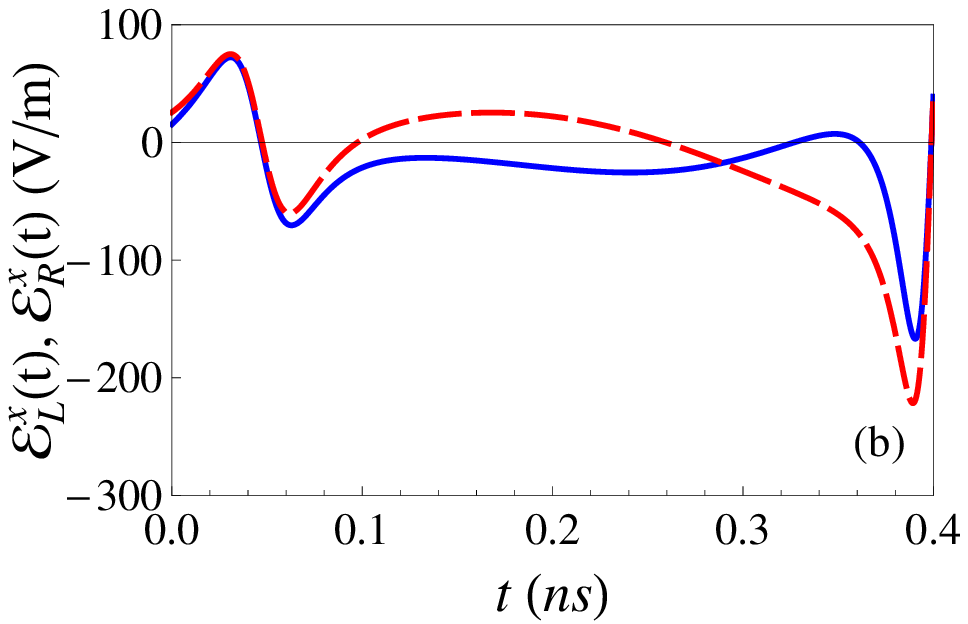}} \scalebox{0.70}[0.70]{%
\includegraphics{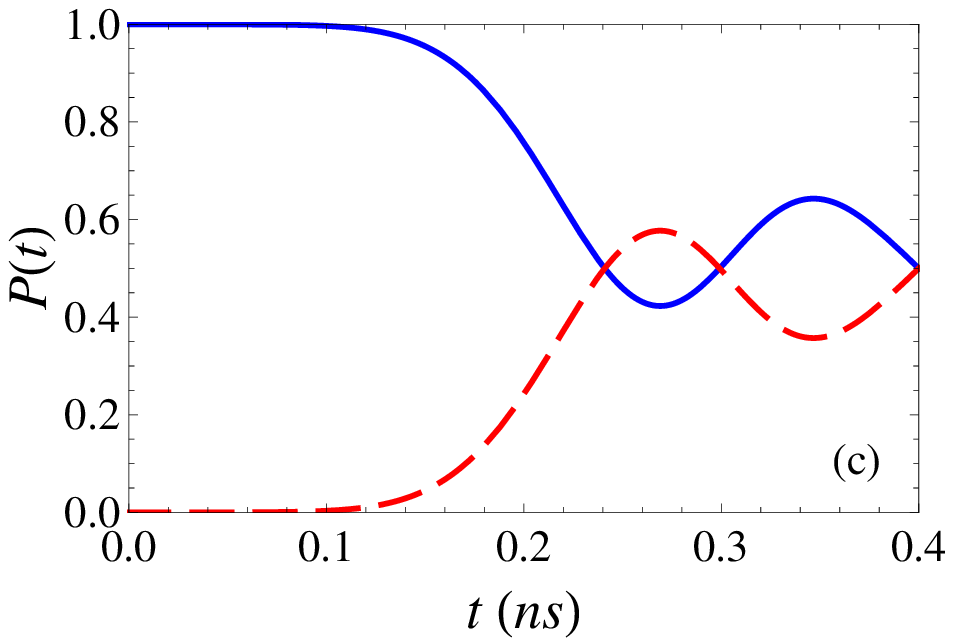}}
\caption{(Color online) Example 3 for two-mode driving with the conditions
at the final state: $\theta_p (t_f)=\pi/2$, $\varphi_p (t_f)=0$, $%
\theta_a(0)=\pi/6$. For $u(t_f)=-2 \pi$: (a) The two auxiliary angles, $%
\theta_a$ (solid blue line) and $\varphi_a$ (dashed red line), compared to
the physical polar angle $\theta_p$ (dotted black line) and the physical
azimuthal angle $\varphi_p$ (dot-dashed orange line). (b) The designed
electric fields ${\mathcal E}^x_L$ (solid blue line) and ${\mathcal E}^x_R$ (dashed red line).
(c) The populations $P_1$ (solid blue line) and $P_2$ (dashed red line).
Other parameters are the same as in Fig. \ref{example1}.}
\label{example3}
\end{figure}
%
We consider now a transition from $\Psi(0)=|1{\rangle}$ to an arbitrary
state at $t=t_f$,
\begin{eqnarray}
\Psi(t_f)= \left(
\begin{array}{c}
\cos{\displaystyle{\frac{\theta_p (t_f)}{2}}} e^{i \varphi_p (t_f)} \\
\sin{\displaystyle{\frac{\theta_p (t_f)}{2}}}
\end{array}
\right).
\end{eqnarray}
Choosing $\varphi_p(t_f)=\varphi_a(t_f)$ for simplicity, the
time-independent coefficients {\ $c_\pm = e^{-i \gamma_\pm(t_f)} {\langle}
\chi_\pm(t_f) | \Psi(t_f) {\rangle}$ are expressed as}
%
\begin{eqnarray}  \label{coefficientc-2}
c_+ &=& e^{-i \gamma_+(t_f)} \cos\frac{\zeta}{2}, \\
c_- &=& e^{-i [\gamma_-(t_f)-\varphi_a(t_f)]} \sin\frac{\zeta}{2},
\end{eqnarray}
where $\zeta=\theta_a(t_f)-\theta_p(t_f)$. As the initial state is $|1 {%
\rangle}$ (up to some phase factor), $P_1(0)$=1 [$P_2(0)=0$] gives the
following condition,
\begin{eqnarray}  \label{P20}
1 - \cos \zeta \cos\theta_a(0) - \sin \zeta \sin \theta_a(0) \cos u(t_f)=0.
\end{eqnarray}

Similarly to the case of state transfer from an arbitrary state to $|2 {%
\rangle}$, Eq. (\ref{P20}) holds when $\cos u(t_f)=\pm 1$. Taking $\cos
u(t_f)=1$ leads to
\begin{eqnarray}
\cos[\zeta - \theta_a(0)] = 1,
\end{eqnarray}
and
\begin{eqnarray}  \label{thetat-example3}
\theta_a(t_f) = \theta_p(t_f) + \theta_a(0).
\end{eqnarray}
We choose the same second-order polynomial \textit{Ansatz} for $\theta_a (t)$
as in Eq. (\ref{thetat-example1}). With a given $\theta_a(0)$, the
coefficients $a_j$ can be solved from the constraint $\dot \theta_a (0) = 0$
and Eq. (\ref{thetat-example3}). Again we set the same form of $\varphi_a
(t) $ as in Eq. (\ref{varphit-example1}). The $b_j$ follow from the
preconditions $\varphi_a(t_f)=\varphi_p(t_f)$, the constraint $\dot
\varphi(0) = (J+\Delta)/\hbar$, and $u(t_f)=2 k \pi$ ($k=0, \pm 1, \pm 2 ...$%
).

In example 3, we show the state transfer from $|1 {\rangle}$ to $%
\Psi(t_f)=(|1 {\rangle} + |2 {\rangle})/\sqrt{2}$, that is, $\theta_p(t_f) =
\pi/2$ and $\varphi_p(t_f)=0$. For $k=-1$, the auxiliary angles $\theta_a$, $%
\varphi_a$ and the physical angles $\theta_p$ and $\varphi_p$ are displayed
in Fig. \ref{example3} (a). This transition realizes a fast Hadamard gate
\cite{Nielsenbook} since $|2 {\rangle}$ becomes $(|1 {\rangle} - |2 {\rangle}%
)/\sqrt{2}$.

\section{Conclusion}

We have studied the possibility to manipulate by electric fields transitions
in a singlet-triplet subspace of a two-electron double quantum dot in the
presence of spin-orbit coupling. By using inverse engineering based on
Lewis-Riesenfeld invariants, we have designed the Hamiltonians that allow
for a fast spin manipulation on a desired timescale. We have applied
two-mode driving using both eigenstates of the dynamical invariant and the
time-dependent difference in their phases. Generally, any fast transition
protocol can be equivalently designed by using a single-mode driving.
However, the two-mode approach is more flexible and provides simpler \textit{%
Ansantzes} for the eigenstates of the dynamical invariants. This technique
has proven useful to minimize the electric fields necessary to perform the
spin operations. This approach can as well be used in other two-level
quantum systems, such as two-level atoms \cite{Chen10b}, Bose-Einstein
condensates in accelerated optical lattices \cite{Oliver}, and
nitrogen-vacancy (NV) center spin in diamond \cite{Zhang}, for engineering
their quantum dynamics.

\section*{Acknowledgement}

We are grateful to A. Ruschhaupt for useful discussions. We acknowledge
funding by the Basque Country Government (Grants Nos. IT472-10 and
BFI-2010-255), Ministerio de Econom\'{i}a y Competitividad (Grant No.
FIS2012-36673-C03-01), UPV/EHU program UFI 11/55, National Natural Science
Foundation of China (Grant No. 61176118), Shanghai Rising-Star Program
(Grant No. 12QH1400800), and Shanghai Pujiang Program (Grant No. 13PJ1403000).

\end{document}